\documentclass{book}
\usepackage[normalem]{ulem}
\usepackage{amsfonts}
\usepackage{amssymb}
\usepackage{amsmath} 					
\usepackage{graphicx} 				
\usepackage{color}           	
\usepackage[ansinew]{inputenc}
\usepackage{comment}					
\begin{document}
\pagestyle{plain}
\pagenumbering{arabic}
\numberwithin{equation}{section}
\numberwithin{figure}{section}
\chapter[Carbon nanotubes: Nonlinear high-Q resonators with strong coupling to single-electron tunneling]{Carbon nanotubes: Nonlinear high-Q resonators with strong coupling to single-electron tunneling\\ {\normalsize Harold B. Meerwaldt, Gary A. Steele, and Herre S.J. van der Zant}}
%
\section{Introduction}
\label{subs:introduction}
A carbon nanotube (CNT) is a remarkable material, which consists of only carbon atoms and can be thought of as a graphene sheet rolled up into a cylinder with the ends capped off with a buckyball sliced in half. This is called a single-walled CNT and, depending on how the graphene sheet is rolled up, CNTs are either semiconducting or metallic. At low temperatures, CNTs contacted by two electrodes become quantum dots, and Coulomb blockade and single-electron tunneling occurs. When more graphene sheets are wrapped up concentrically a multiwalled CNT is formed.

CNTs are fabricated using various methods. These include arc discharge \cite{Journet1997}, laser ablation \cite{Thess1996}, chemical vapor deposition (CVD) \cite{Kong1998}, and the use of high pressure carbon monoxide \cite{Nikolaev1999}. There are two ways to make doubly-clamped CNT resonators. The first method starts with growing CNTs on an oxidized silicon wafer and then underetching them with a trench so that they become suspended. This method exposes the CNT to resist, the electron beam, and acid, and may give rise to defects in and residues on the CNT. In the ultraclean method \cite{Cao2005}, processing on the CNT is avoided by first making the contacts and the trench and in the final step growing the CNTs. A suspended CNT resonator can be fabricated without defects, thus reducing damping which results in quality factors for the flexural modes above 100,000 \cite{Huettel2009}. 

Having a radius that can be as small as a nanometer and lengths of several micrometers, CNTs have aspect ratios in the thousands, on a scale and purity that is difficult to achieve in resonators fabricated top-down from silicon, SiN, or other materials. The enormous aspect ratio makes it easy to excite the CNT such that the displacement of the flexural mode is similar in magnitude to the radius; the CNT thus behaves as a thin, narrow beam resonator. The combination of low damping and high aspect ratio paves the way for nonlinear effects. 

There are two other remarkable properties of CNTs: Their Young's modulus of 1.3 TPa makes them an incredibly stiff material, and combining this with a density of only 1350 $\textrm{kg/m}^3$, CNT resonators reach frequencies of several hundreds of megahertz and even gigahertz. At millikelvin temperatures, with $\hbar \omega_0 \gg k_b T$, thermal phonons can no longer excite the flexural motion. This makes CNTs perfect candidates to observe quantum (nonlinear) effects in mechanical resonators \cite{Dykman1988}.

This chapter is built up as follows. In the next section various ways of detecting the motion of CNT resonators are discussed. In the third section we show how single-electron tunneling gives rise to frequency softening and damping. Section four contains Duffing-like nonlinearities caused by geometry, electrostatics, and single-electron tunneling. In section five we look at nonlinearities from parametric driving and mode coupling to other dimensions of oscillation and to higher modes. Table \ref{tab:prop} gives an overview of the symbols used in this chapter for variables and constants along with typical values and example device parameters.
\begin{table*}[h]
	\centering
		\begin{tabular}{|c c c|}
		\multicolumn{3}{c}{\textbf{General properties of single-walled CNTs}}\\
		\hline
			$E$&Young's modulus& 1.3 TPa\\
			$\rho$&mass density&1350 kg/m$^3$\\
			\hline
		 \multicolumn{3}{c}{}\\ 
		 \multicolumn{3}{c}{\textbf{Example CNT}}\\ 
		 \hline
		 $L$&length&800 nm\\
		 $r$& radius&1.5 nm\\
		 $h_g$&distance to gate&230 nm\\
		 $C_g$&capacitance to gate&10 aF\\
		 $C_{tot}$ &total capacitance&20 aF\\
		 $Q$&quality factor&100,000\\
		 $\omega_0/2\pi$&resonance frequency&140 MHz\\
		 $I$&second moment of inertia& 4.0 nm$^4$\\
		 $A$&cross-sectional area&7.1 nm$^2$\\
		 \hline
		 \multicolumn{3}{c}{}\\ 
		 \multicolumn{3}{c}{\textbf{Variables}}\\ 
		 \hline
		  $u(z,t)$&displacement&-\\
		 $\mathcal{A}(t)$&slowly-changing amplitude&-\\
		 $\xi(z)$&mode shape&-\\
			\hline

		 \end{tabular}
		\caption{General properties of a single-walled CNT, device parameters for a CNT used as an example in the subsequent sections, and variables.}
		\label{tab:prop}
\end{table*}

\section{Detecting the motion of a carbon nanotube resonator}
\label{sec:2}

In this section, we discuss different ways to detect the flexural motion of CNT resonators. Other vibrational modes such as breathing and stretching modes are also present in CNT resonators, but non-flexural modes are difficult to actuate and detect. They are therefore not discussed in this chapter. Because a CNT is easily perturbed, detection has a large influence on its motion. When discussing the nonlinear dynamics of CNT resonators it is therefore important to have an overview of the different detection methods. Optical detection, which is used frequently in top-down devices, is not an option because of the small cross-section of a CNT. Until now, detection has therefore been done through microscopy or electronically. We summarize five different methods to observe its motion: scanning force microscopy, transmission electron microscopy, field emission microscopy, the mixing technique, which involves frequency down-mixing the flexural motion from several hundreds of megahertz to several kilohertz, and the rectification technique, where the amplitude of the flexural motion is obtained by measuring the dc current flowing through the CNT. The last two (electronic) methods are self-detecting because the CNT is both the object studied and the detector.

\begin{figure}[h]
\centering
	\includegraphics[scale=0.50]{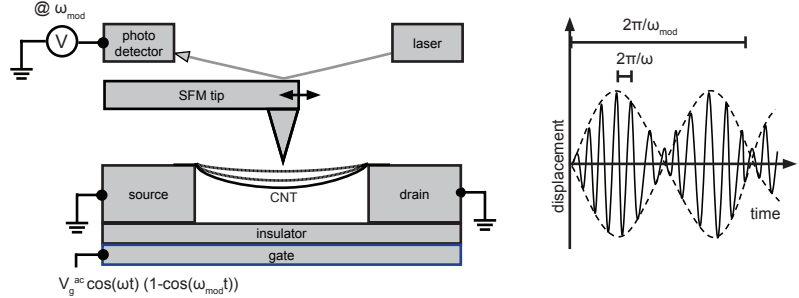}
	\caption{Scanning force microscopy. A suspended doubly-clamped CNT is brought into motion by a nearby gate electrode (left). The amplitude-modulated motion (right) at different parts of the CNT is read out at $\omega_{mod}$ by a tip in tapping mode.}
	\label{fig:bandwSetupSFM}
\end{figure}

\begin{figure}[h]
\centering
	\includegraphics[scale=0.60]{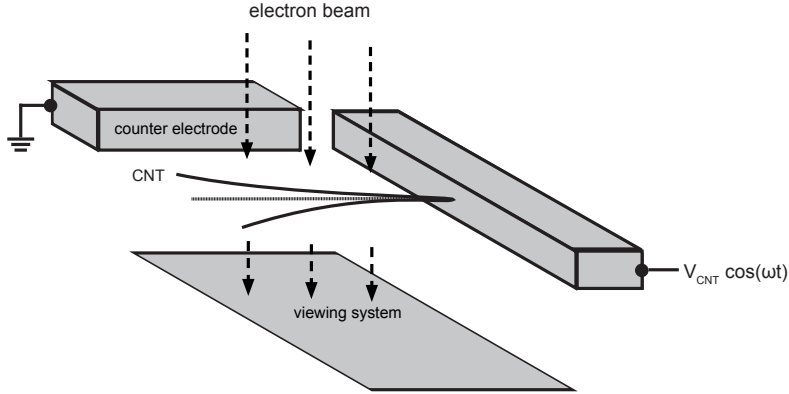}
	\caption{Transmission electron microscopy. A suspended CNT cantilever is driven by applying a voltage $V_{CNT}$ to the CNT with respect to a grounded counter electrode. Electrons from an incident electron beam are not transmitted through the CNT and as result an image of the flexural motion is obtained.}
	\label{fig:bandwSetupTEM}
\end{figure}

Using scanning force microscopy (SFM) \cite{Garcia-Sanchez2007}, it is possible to spatially image the different mode shapes in a CNT resonator. Figure \ref{fig:bandwSetupSFM} shows an SFM in tapping mode with the tip positioned above a doubly clamped CNT. The CNT is actuated by a nearby gate at a frequency $\omega$. Since the bandwidth of the SFM is not high enough to measure the oscillations of the CNT at the resonance frequency, an amplitude-modulated actuation signal is used for the voltage on the gate with a frequency of $\omega_{mod}$. The tip cannot follow the high-frequency CNT vibrations, but it can follow the envelope of the oscillation. On resonance, the amplitude-modulated actuation results in a fast-oscillating displacement envelope $u$ with frequency $\omega_{mod}$ (see figure \ref{fig:bandwSetupSFM}). The modulation frequency can be conveniently chosen such that the signal can be measured with a lock-in amplifier. For higher sensitivity, $\omega_{mod}$ is matched to the first eigenmode of the SFM tip. The amplitude of oscillation at different locations on the CNT provides a time averaged image of the shape of the driven mode. 
With transmission electron microscopy (TEM) \cite{Poncharal1999}, a fiber of CNT cantilevers is connected to a gold wire, which is placed between the electron gun and the viewing system, as shown in figure \ref{fig:bandwSetupTEM}. By applying a voltage to the CNTs, the ones that are not perpendicular to the grounded counter electrode are attracted to it. To drive the CNTs, an alternating voltage is applied to them and the resulting envelope of the motion is observed in the TEM image. As with the SFM method, the low bandwidth of TEM imaging allows visualizing only of the mode shape of the vibrations.

Reports on these two methods have not given indications of nonlinearity. In the experiment using SFM quality factors in the order of 10 and the disturbance due to the tapping tip may have prevented nonlinearity from being observed. In the experiment with TEM, the high aspect ratio of the CNT makes it difficult for the CNT cantilever to bend to such an extent that geometric nonlinearity becomes relevant \cite{Venstra2010}. Even though the amplitude of oscillation is far larger than the CNT radius, tension induced by bending is much smaller in CNT cantilevers than in doubly-clamped CNTs.

\begin{figure}[h]
\centering
	\includegraphics[scale=0.60]{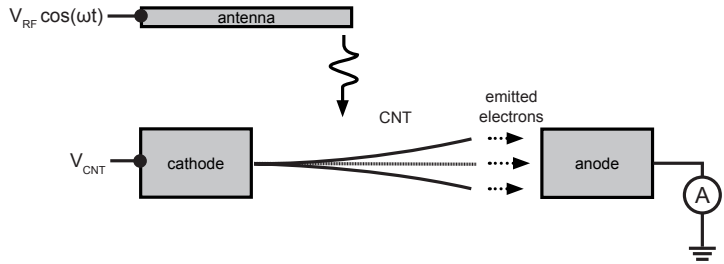}
	\caption{Field emission microscopy. A suspended CNT cantilever is excited by radio waves from a nearby antenna. Current tunnels from the CNT to the anode and oscillates with the motion of the CNT resonator.}
	\label{fig:bandwSetupFEM}
\end{figure}

With field-emission microscopy (FEM), temporal and spatial information on the flexural motion is obtained. Figure \ref{fig:bandwSetupFEM} shows a CNT cantilever to which a few hundred volts is applied. A field-emission tunnel current now flows to the anode which is placed up to a few hundred nm away. In one case \cite{Jensen2007}, the CNT was excited by radio waves from a nearby antenna. At the mechanical resonance frequency the distance between the tip of the CNT, which acts as a cathode, and the anode oscillates. This causes the tunnel resistance and therefore the current flowing through the circuit to oscillate. In this way, the CNT acts as a rectifier; the current is highest when the CNT is in its equilibrium position and becomes lower, but never negative, when the CNT moves away from this position in either direction. As a consequence, the time-averaged current decreases when the CNT is set in motion. Thus, the ac motion of the the CNT is probed by a dc measurement of the averaged current. In another case \cite{Perisanu2010} the CNT was positioned near a phosphor screen so that the averaged 2-dimensional image of emitted electrons was recorded by a 25 Hz video camera. 

It should be noted that the three techniques discussed up to now operate at room temperature. At low temperatures, mechanical properties such as damping change dramatically. We now discuss two electronic methods that are used at low temperature, which we refer to as the mixing technique and the rectification technique.

\begin{figure}[h]
\centering
	\includegraphics[scale=0.60]{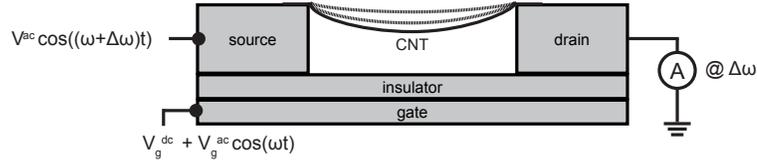}
	\caption{Detecting the CNT's motion using the mixing technique. A CNT is doubly clamped between a source and a drain electrode suspended above and driven by a gate electrode. The flexural motion of the CNT mixes the bias voltage at $\omega +\Delta \omega$ and the gate voltage at $\omega$ to a current at $\Delta \omega$. When frequency-modulation of the source voltage is performed, no ac voltage needs to be applied to the gate electrode.}
	\label{fig:bandwSetupmixing}
\end{figure}

In measuring the high-frequency response of mechanical resonators, one often employs electrical frequency-mixing techniques. Here, a nonlinear element mixes two different high-frequency signals to their difference frequency, which is considerably lower and is measured more easily. In the case of a CNT mechanical resonator, the CNT itself is used as the nonlinear element. To measure CNTs using the mixing technique \cite{Sazonova2004,Peng2006,Witkamp2006,Lassagne2008}, the working principle is modeled as follows. A three terminal setup is used, as shown in figure \ref{fig:bandwSetupmixing}. The current $I=GV$ flowing through the CNT, where $G$ is the conductance of the CNT and $V$ is the bias voltage over the CNT, depends on the gate induced charge $q_c=C_gV_g$, where $C_g$ is the capacitance to the gate and $V_g$ is the gate voltage. When the CNT moves away from its equilibrium position, the capacitance to the gate changes, giving $q_c^{mech}=C_g^{ac}(t)V_g$. Also, when an oscillating signal $V_g^{ac}(t)$ is applied to the gate, the gate induced charge changes as: $q_c^{direct}=C_gV_g^{ac}(t)$. Putting these two contributions together gives the time-dependent conductance:
\begin{equation}
G^{ac}(t)=\frac{d G}{d V_g}\left(V_g^{ac}(t)+\frac{C_g^{ac}(t)}{C_g}V_g\right).
\label{eq:mixing}
\end{equation}
The two contributions both oscillate at a frequency $\omega$ but can in general have a different phase. Note that the frequency of the mechanical oscillation of the CNT equals the frequency of the driving voltage applied to the gate electrode. When also the bias voltage is alternated, but now at a frequency $\omega+\Delta \omega$, both the mechanical oscillations and the direct oscillations are mixed given that $I=G^{ac}(t)V^{ac}(t)$. One mixing component occurs at $\Delta \omega$, which can be conveniently chosen at a few kHz. This signal is then detected using a lock-in amplifier. At high driving, nonlinear effects become visible \cite{Sazonova2004,Witkamp2006}.

Measurement of the flexural motion of a CNT resonator has also been performed using the mixing technique with a frequency-modulated voltage applied to the source electrode \cite{Gouttenoire2010}. The frequency modulation of the source voltage causes sidebands in the spectrum around the drive frequency. They both mix with the mechanical resonance and result in a signal at the modulation frequency. In this case no signal needs to be applied to the gate as the CNT is actuated by setting the carrier frequency of the source voltage to the mechanical resonance frequency. 

Another technique that resembles the mixing technique transduces the mechanical oscillations of the CNT resonator at several hundred megahertz into a dc current \cite{Huettel2009}, in this way using the CNT resonator as a rectifier. The advantage of this technique is that the amplitude of the mechanical motion is given by a change in the dc current flowing through the CNT. As with the mixing technique, the use of the rectification technique avoids the difficulty of getting small, high-frequency signals out of a setup at millikelvin temperatures and minimizing the cross-talk from the actuation onto the measurement signal.

With the rectification technique, the working principle is as follows. The CNT is suspended between source and drain electrodes above a gate electrode. At low temperatures, the CNT acts as a suspended quantum dot in which charging effects dominate transport. It is actuated using a nearby antenna, which sends out an oscillating electric field. When the CNT oscillates the distance to the gate changes, and therefore the capacitance $C_g^{ac}(t)=\frac{d C_g}{d u}u_{ac} \cos(\omega t)$, where $u$ is the displacement of the CNT. Since the only relevant quantity is the charge induced by the gate, it is equivalent to say that the gate voltage changes as $V_g^{ac}(t) = \frac{C_g^{ac}(t)}{C_g}V_g$. Expanding the current through the CNT quantum dot with respect to gate voltage one obtains:
\begin{eqnarray}
I(V_g+V_g^{ac}(t))=I(V_g)+\frac{d I}{d V_g}\frac{V_g}{C_g}\frac{d C_g}{d u}u_{ac} \cos(\omega t)\nonumber\\
+\frac{1}{2}\frac{d^2 I}{d V_g^2}\left(\frac{V_g}{C_g}\frac{d C_g}{d u}\right)^2u_{ac}^2 \left(\frac{1}{2}+\frac{1}{2}\cos(2\omega t)\right).
\label{eq:}
\end{eqnarray}
The high resistance of CNTs combined with a high capacitance to ground often means that the bandwidth is too low to measure the oscillations at $\omega$ and $2\omega$ directly. Omitting these oscillating terms gives for the change in the dc current $\Delta I$ due to mechanical motion:
\begin{equation}
\Delta I= \frac{1}{4}\frac{d^2 I}{d V_g^2}\left(\frac{V_g}{C_g}\frac{d C_g}{d u}\right)^2u_{ac}^2.
\label{eq:deltaI}
\end{equation}
Important to note is that the change in current is averaged over time and that it is proportional to the amplitude squared. 

Equation \ref{eq:deltaI} shows that, in order to observe a large mechanical signal, it is required that $d^2I/d V_g^2$ be high, as is the case in the Coulomb blockade regime \cite{Beenakker1991,Thijssen2008}. When measured on a Coulomb peak the second derivative is negative and therefore the change in current as well. In the Coulomb valley both the second derivative and the change in current are positive. Typically, $d^2 I/d V_g^2$ is several $\mu A/V^2$. On the inflection point of the Coulomb peak $d^2I/d V_g^2=0$ and there is no change in current. This is in contrast to the mixing technique which, as shown in equation \ref{eq:mixing}, has the highest signal on the inflection point and the lowest on the Coulomb peak or in the valleys.
\begin{figure}[h]
\centering
	\includegraphics[scale=0.60]{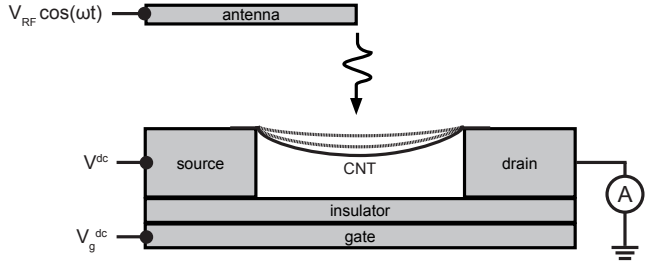}
	\caption{Detecting the CNT's motion using the rectification technique. A CNT is doubly clamped between a source and a drain electrode suspended above a gate electrode. The CNT is driven by a high-frequency signal applied to an antenna at approximately 1 cm from the CNT. The single-electron tunnel current through the CNT quantum dot contains information on the oscillation of the CNT at dc due to a rectification effect from Coulomb blockade.}
\end{figure}

\section{Linear dynamics of a carbon nanotube resonator strongly coupled to a quantum dot}

Because of their high quality factor and therefore narrow linewidths, ultraclean CNT resonators allow for the detection of small changes in resonance frequency. In this section we describe the linear response of CNT resonators measured with the rectification technique. These measurements are performed at low temperatures; the CNT then acts as a quantum dot \cite{Tans1997,Bockrath1997,Cobden2002}, and single-electron tunneling is the dominant transport mechanism. We consider the backaction force caused by tunneling electrons and how this stochastic force gives rise to dips in the resonance frequency and increased damping. 

\begin{figure}
\centering
	\includegraphics[scale=0.60]{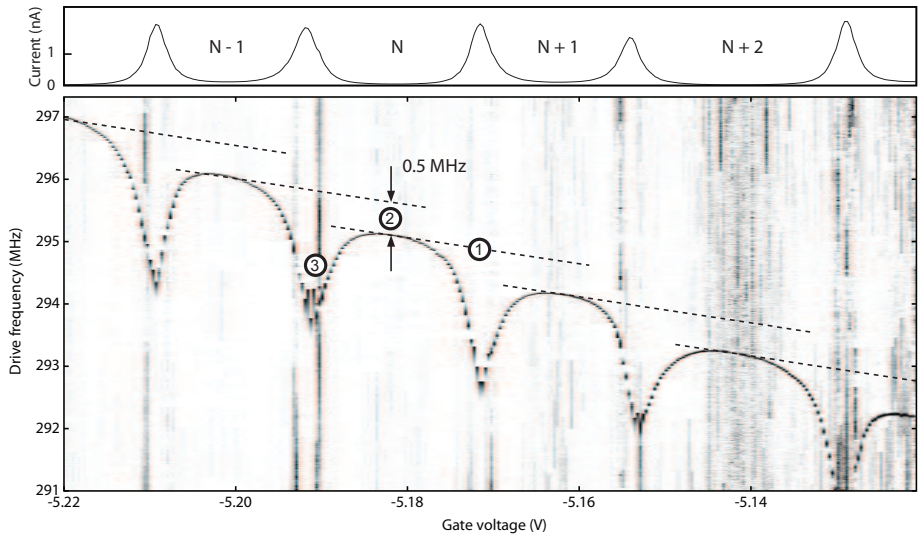}
	\caption{Top: Coulomb oscillations of the current through a CNT quantum dot; in Coulomb blockade current is zero and the indicated (negative) number of electrons on the CNT is fixed. Between charge states a single-electron tunneling current is visible as Coulomb peaks. Bottom: Gate dependence of resonance frequency (from \protect\cite{Steele2009}); the features  are indicated by the encircled numbers: (1) a slope caused by tension at fixed charge, (2) a frequency offset caused by a tension difference between charge states, and (3) frequency dips caused by single-electron tunneling.}
	\label{fig:freqdips}
\end{figure}

Figure \ref{fig:freqdips} shows Coulomb oscillations (top), and the change in resonance frequency with gate voltage (bottom) at negative gate voltages. Three features are worth mentioning: i) in Coulomb blockade, there is a fixed charge on the CNT and the resonance frequency-gate voltage curve approaches a linear dependence, denoted by the dashed lines; ii) from one charge state to the next, there is an offset between the sloped lines; and iii) at the position of Coulomb peaks corresponding to a transition between charge states, there are resonance frequency dips arising from a softening-spring behaviour.

The slope in resonance frequency is explained as follows: in Coulomb blockade the charge on the CNT quantum dot is constant. If a more negative gate voltage is applied, the force acting on the CNT increases, which pulls the CNT more towards the gate. The tension caused by this pulling leads to the first feature: the resonance frequency increases with gate voltage.

As the gate voltage becomes more negative, more and more electrons are extracted from the CNT quantum dot, leaving behind holes. From one charge state to the next, there is a difference in charge of one electron. The tension arising from an extra electron causes the resonance frequency to change in a discrete manner, leading to the second feature: an offset in resonance frequency between two charge states. Figure \ref{fig:freqdips} shows a frequency offset of $0.5$ MHz, which is more than 100 times the linewidth. 

\begin{figure}
\centering
	\includegraphics[scale=0.80]{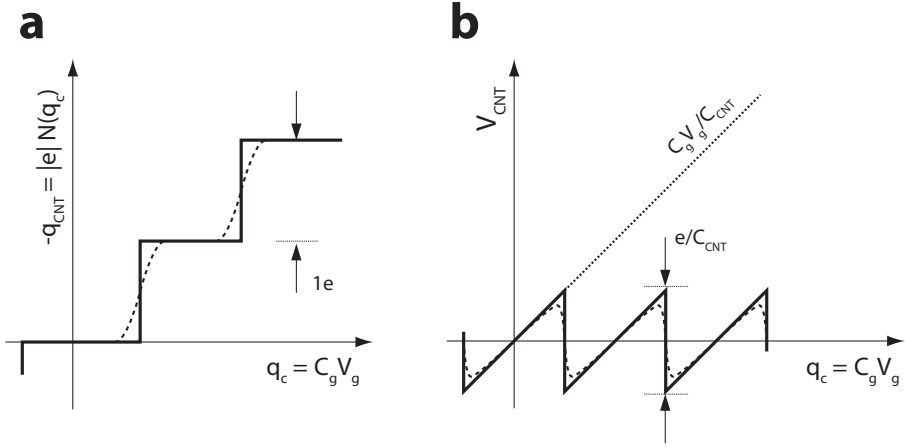}
	\caption{Left: The charge on the CNT increases with integer steps as the gate induced charge is increased (solid line). At finite temperature or finite tunnel coupling the charge transitions are smoothed out (dashed line). Right: The voltage on the CNT increases with increased gate induced charge as $V_{CNT}=C_gV_g/C_{CNT}$ until an extra charge enters the CNT, at which point the CNT voltage decreases with $e/C_{CNT}$ (solid line). At finite temperature or finite tunnel coupling this voltage jump is smoothed again (dashed line).}
	\label{fig:Screening}
\end{figure}

To explain the frequency dips arising from single-electron tunneling, we look at the electrostatic force acting on the CNT:
\begin{equation}
F_{dc}=-\frac{d U}{d u} =\frac{1}{2} \frac{d C_g}{d u} (V_g-V_{CNT})^2,
\label{eq:SETfor}
\end{equation}
where $d C_g/d u$ is the derivative of the capacitance to the gate with respect to displacement and $V_g$ is the gate voltage. The voltage on the CNT, $V_{CNT}$, is given by:
\begin{equation}
	V_{CNT}=\frac{C_gV_g+q_{CNT}}{C_{CNT}},
\end{equation}
where $q_{CNT}$ is the charge on the CNT and $C_{CNT}=C_g+C_s+C_d$ is the total capacitance of the CNT to gate, source, and drain. The charge on the CNT is given by:
\begin{equation}
q_{CNT} = -eN(q_c),
\label{eq:}
\end{equation}
where $e$ is the elementary charge, and $N(q_c)$ is the number of electrons on the CNT, which is negative in figure \ref{fig:freqdips}. The charge and voltage on the CNT are shown in figure \ref{fig:Screening}. The gate induced charge $q_c$ is the charge that would be on the CNT if there were no Coulomb blockade, and in the case of antisymmetrically applied voltages ($V_s=-V_d$) and equal capacitances ($C_s=C_d$) for source (s) and drain (d), it is given by:
\begin{equation}
	q_c=C_gV_g.
\end{equation}
As the CNT undergoes a charge transition, electrons can hop on and off the CNT quantum dot. Whereas in Coulomb blockade there was no possibility to screen the voltage on the gate, in a charge transition the CNT overscreens the gate voltage, so that on average the voltage on the CNT remains zero. It is this overscreening which causes a reduction in tension and gives rise to the third feature: dips in the resonance frequency due to single-electron tunneling. Figure \ref{fig:freqdips} shows that single-electron tunneling can lead to a dip in the resonance frequency by as much as 2 MHz. The change in resonant frequency due to single-electron tunnelling over this small gate range is an order of magnitude larger than that from the gate induced mechanical tension, which is described in section \ref{subs:geom}.

By considering the electrostatic force, the change in spring constant and thereby the change in resonance frequency due to single-electron tunneling can be calculated:
\begin{eqnarray}
\Delta k_{SET} = - \frac{d F_{CNT}}{d u}&\approx&(V_g-V_{dot})\frac{d C_g}{d u}\frac{d V_{CNT}}{d u}\nonumber \\
&\approx &\frac{V_g(V_g - V_{CNT})}{C_{CNT}}\left(\frac{d C_g}{d u}\right)^2\left(1-e\frac{d \left\langle N \right\rangle}{d q_c}\right),
\label{eq:deltak}
\end{eqnarray}
where we neglect slowly varying terms with $d ^2C_g/d u^2$.

The average number of electrons, or average occupancy $\langle N\rangle$, can be non-integer (dashed line in figure \ref{fig:Screening}). At fixed $-q_{CNT}$, $d \langle N\rangle/d q_c$ equals zero and the factor $V_g(V_g-V_{CNT})$ represents the tension causing the first two features, the frequency slope and the frequency offset. At a charge transition, $d \langle N\rangle/d q_c$ is positive, and this gives rise to softening-spring behaviour in equation \ref{eq:deltak}. The sharper the charge transition is, the deeper the frequency dip is. This follows from the fact that when $d\langle N\rangle/dq_c$ is higher, the average number of charges on the CNT changes more as the CNT oscillates towards and away from the gate, in this way giving rise to a larger change in force with change in displacement, $dF_{CNT}/du$, and therefore a larger change in spring constant.

\begin{figure}[h]
\centering
	\includegraphics[scale=0.90]{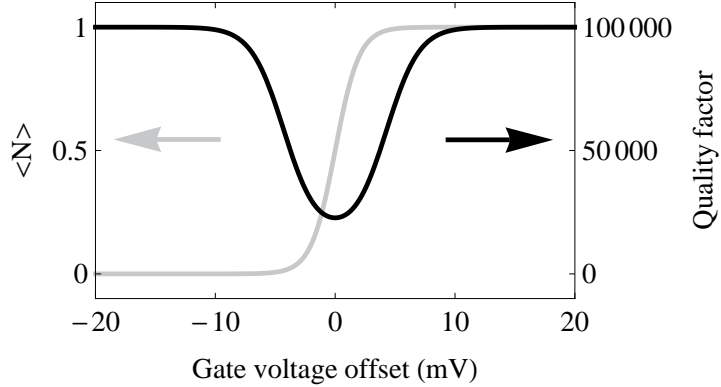}
	\caption{Noninteger part of the average occupancy (left axis) at a charge transition from $N$ to $N+1$ electrons and the corresponding change in quality factor (right axis) through single electron tunneling for the example CNT with 20 electrons at $V_g=3$V assuming $\Gamma_{tot}=10^{13}\textrm{Hz}$. The maximum of the Coulomb peak is located at 0 mV of gate voltage offset.}
	\label{fig:Qfactor}
\end{figure}

\begin{figure}[h]
\centering
	\includegraphics[scale=0.70]{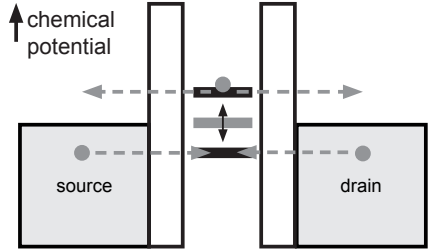}
	\caption{Energy diagram (exaggerated): Due to the asymmetry imposed by the Fermi level of the source and drain, electrons are allowed to tunnel onto the CNT but not off the CNT when the chemical potential of the CNT is lower than the Fermi level of the leads. When the chemical potential of the CNT is higher than the Fermi level of the leads, electrons are allowed to tunnel off the CNT but there are no electrons to tunnel onto the CNT.}
	\label{fig:energydiagramdamping}
\end{figure}

Using $\Delta k=2m\omega_0\Delta\omega_0$ we arrive at the change in resonance frequency due to single-electron tunneling:
\begin{equation}
	\Delta \omega_0^{SET}=\frac{V_g(V_g - V_{CNT})}{2m\omega_0 C_{CNT}}\left(\frac{dC_g}{du}\right)^2\left(1-e\frac{d\left\langle N \right\rangle}{dq_c}\right).
\end{equation}
In addition to a change in resonance frequency, electrons hopping on and off the CNT quantum dot can also cause damping of the mechanical motion. We look at the limit when there are many tunneling events per mechanical oscillation, $\Gamma_{tot} \gg \omega_0$. As derived in \cite{Labadze2011}, the total damping has an intrinsic contribution and a contribution due to the stochastic backaction force associated with tunneling electrons:
\begin{equation}
\frac{\omega_0}{Q_{tot}}=\frac{\omega_0}{Q_{int}}+\frac{F_{stoch}}{m}\frac{1}{\Gamma_{tot}}\frac{d}{d u}\left (\frac{\Gamma^+}{\Gamma_{tot}}\right ),
\label{eq:}
\end{equation}
where $Q_{tot}$ is the total quality factor, $Q_{int}$ is the quality factor in the absence of tunneling electrons, $\Gamma^\pm$ is the rate of electrons tunneling onto (off) the CNT, and $\Gamma_{tot}=\Gamma^++\Gamma^-$ is the total tunneling rate. The stochastic force the CNT experiences, $F_{stoch}=F(N+1)-F(N)$, is the difference between the force experienced at $N$ and $N+1$ electrons. Assuming that both source and drain voltage are much smaller than the gate voltage, $V_s, V_d\ll V_g^{dc}$, and taking only into account the dc force experienced by the electrons as the CNT oscillates (which is valid as long as $V_g^{ac}\ll V_g^{dc}$), reference \cite{Labadze2011} gives the following expression for the stochastic force:
\begin{equation}
	F_{stoch}=\frac{1}{C_{tot}^2}\frac{d C_g}{d u}\left(2e(C_L+C_R) V_g+e^2(2N+1)\right).
	\label{eq:Fstoch}
\end{equation}
In this limit, the stochastic force on the CNT depends on the distance of the CNT to the gate and consequently on the average number of charges on the CNT, both of which change as the CNT oscillates. To obtain $F_{stoch}$ for holes, the substitution $e\rightarrow-e$ must be made in equation \ref{eq:Fstoch} with $N$ now the number of holes. Using 
\begin{equation}
\left\langle N \right\rangle = \frac{1/\Gamma_-}{1/\Gamma_++1/\Gamma_-}=\frac{\Gamma_+}{\Gamma_{tot}},
\end{equation}
this leads to the following expression for the quality factor:
\footnotesize
\begin{eqnarray}
Q_{tot}=\left(\frac{1}{Q_{int}}+ \frac{V_g^{dc}}{m\omega_0\Gamma_{tot}C_{tot}^2C_g}\left(\frac{d C_g}{d u}\right)^2\left(2e(C_L+C_R)V_g^{dc}+e^2(2N+1)\right)\frac{d\left \langle N\right\rangle}{dV_g}\right)^{-1}.
\label{eq:}
\end{eqnarray}
\normalsize
Figure \ref{fig:Qfactor} illustrates how a charge transition from $N$ to $N+1$ gives rise to a decrease in the quality factor. Similar to the frequency dips discussed earlier, damping is highest when $d\left \langle N\right\rangle/dV_g$ is highest. At this point, the mechanical oscillation causes the largest change in the average occupancy and consequently the influence of the backaction is the most pronounced.
Physically the damping is caused by energy dependent tunneling. When the CNT is oscillating, the change in capacitance to the gate makes the chemical potential of the CNT oscillate. Even though many electrons tunnel onto and off the CNT during one mechanical oscillation, on average more tunnel on when the energy level is lower, and more tunnel off when the energy level is higher. This asymmetry is imposed by the Fermi energy of the source and drain. Figure \ref{fig:energydiagramdamping} shows that, when the chemical potential of the CNT is lower than the Fermi level of the source or drain, electrons are allowed to tunnel onto the CNT but not off the CNT. Conversely, when the chemical potential of the CNT is higher than the Fermi level of the source or drain, electrons are allowed to tunnel off but there are no electrons to tunnel onto the CNT. Because of this asymmetry, electrons will tunnel on when the level is pulled below the Fermi energy by the motion, and tunnel out when the level is pushed above the Fermi energy by the motion, thereby extracting energy from the mechanical mode, damping it's motion. Note that this energy is carried away by hot electrons above the Fermi energy in the leads, where it is later dissipated when these electrons relax.
We finally note that the gate-dependent damping, as illustrated by the black curve in figure \ref{fig:Qfactor}, has been observed in experiments. The change in Q can be dramatic: from $10^5$ in the Coulomb valleys to several hundreds near the Coulomb peak. At present we are performing a more accurate comparison with the theoretical calculations \cite{Meerwaldt2012}.


\section{Carbon nanotube resonators in the nonlinear regime}
\label{sec4}
Hooke's Law states that, when an elastic object is brought out of its equilibrium position, the restoring force is linearly dependent on the displacement. As the applied force and resulting displacement increase, Hooke's Law is no longer valid and nonlinear effects become important. The first correction to Hooke's Law is made by adding a restoring force term that is quadratic in displacement. The quadratic term leads to a renormalization of the cubic term \cite{Lifshitz2008}. In many cases, this renormalization is a small effect, and hence we have neglected it in this analysis. In this section we focus on the next term of the restoring force, which is cubic in displacement. The cubic term is called the Duffing term, and the resulting equation of motion the Duffing equation (see also \cite{Lifshitz2008}). The restoring force is now given by $F_r=-k_0u-m\alpha u^3$, were $\alpha$ is the nonlinearity or Duffing parameter with units of N/(kg m$^3$). The nonlinearity can also be seen as a correction to the spring constant. Writing the restoring force as $F_r=-ku=-(k_0+m\alpha u^2)u$, the nonlinearity parameter is given by:
\begin{equation}
\alpha=\frac{1}{2m}\frac{d^2 k}{du^2}
\label{eq:}
\end{equation}

When $\alpha$ is positive, the Duffing term causes, as drive force is increased, the resonance peak to tilt towards higher  frequencies, resulting in a stiffening-spring behaviour. Conversely, when $\alpha$ is negative and the resonance peak tilts towards lower frequencies, it results in a softening-spring behaviour. As the applied force is increased further, the tilting of the peak causes bistability: for one drive frequency there are two metastable states of oscillation, one with a low amplitude and the other with a high amplitude of oscillation \cite{Dykman1988}.

In this section we first describe how the measured lineshape changes as the drive power is increased and at what power the lineshape becomes nonlinear. Next, we describe three causes for nonlinearity in CNT mechanical resonators: geometric nonlinearity caused by mechanical tension, electrostatic nonlinearity, and nonlinearity due to single-electron tunneling. We calculate the corresponding $\alpha$ values for the example CNT described in table \ref{tab:prop}.

\subsection{Nonlinear response of carbon nanotube resonators}
\begin{figure}[h]
	\centering
	\includegraphics[scale=0.50]{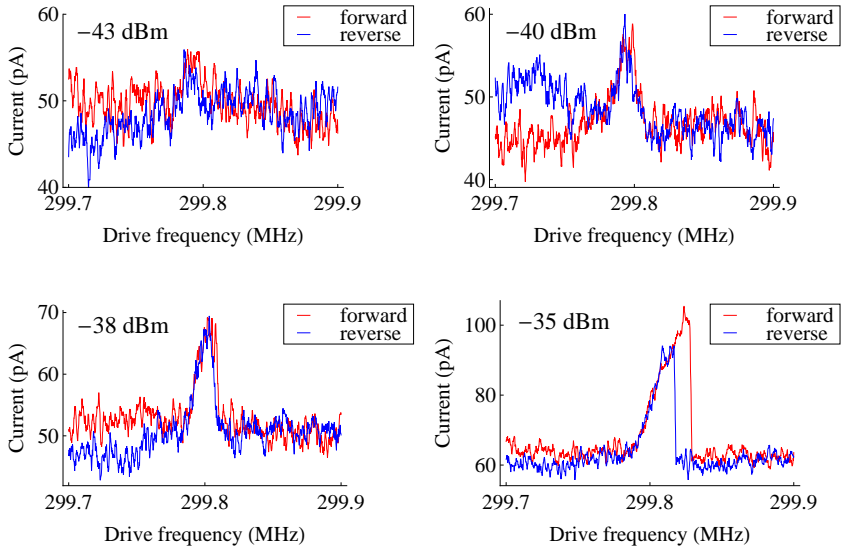}
	\caption{Frequency responses of the rectified current (see previous section) showing the mechanical signal drowned in noise (-43 dBm), a linear response with a Lorentzian lineshape (-40 dBm), the onset of nonlinearity (-38 dBm), and nonlinearity with bistability and hysteresis (-35 dBm), of a CNT across a 600 nm wide trench, 285 nm above a gate, measured at 20 mK.}
	\label{fig:powerdependence}
\end{figure}

Figure \ref{fig:powerdependence} shows a measurement of the lineshape of a 600 nm long CNT resonator detected by using the rectification technique (see section \ref{sec:2}). At a low power of \linebreak -43 dBm, the CNT is driven to oscillate at a small amplitude and the noise is too high to observe the motion. As the drive power is increased to -40 dBm, a Lorentzian lineshape appears in the current when the drive frequency is near the mechanical resonance frequency of the CNT. At this drive power, the restoring force is given by Hooke's Law and can be approximated to be linear with displacement. When the drive power is increased further to -38 dBm, the restoring force is nonlinear with respect to displacement but the amplitude is still single-valued, resulting in a sharkfin-like appearance of the lineshape. At a higher power of -35 dBm in figure \ref{fig:powerdependence}, bistability and hysteresis occur. Bistability appears as for certain frequencies there is both a low- and a high-amplitude state in which the resonator can oscillate. The amplitude of the motion depends on the direction in which the drive frequency is changed, displaying a hysteretic behaviour when the drive frequency is swept from high to low, or from low to high. The bottom two panels of figure \ref{fig:powerdependence} show nonlinearity in the form of a Duffing sharkfin-like lineshape. The sharkfin lineshape points towards higher frequency, indicating a stiffening spring behaviour.
In many applications \cite{Craighead2000,Ekinci2005}, the range of powers over which the mechanical resonance displays a detectable linear response is an important parameter. This range of powers is often referred to as the ``dynamic range'' of the mechanical oscillator. The smallest useful drive power is the one for which the measured signal matches the measured noise, and the largest useful drive power is the one at the onset of nonlinearity. 

To estimate an upper bound for the dynamic range, we follow the analysis of \cite{Postma2005} and only consider the thermomechanical contribution to the noise, leaving out other intrinsic and extrinsic sources of noise. The spectral density of displacement noise at the resonance frequency is:
\begin{equation}
	S_u=\frac{4 k_bTQ}{m\omega_0^3},
\end{equation}
where $k_b$ is Boltzmann's constant, $T$ is temperature of the resonator, $Q$ is the quality factor, $m$ is the total mass of the resonator, and $\omega_0$ is the resonance frequency. For the example CNT in table \ref{tab:prop} this gives a displacement noise of $\sqrt{S_u}=0.15 \textrm{ pm}/\sqrt{\textrm{Hz}}$.
The root-mean-square amplitude of oscillation at which nonlinearity sets in, $\mathcal{A}_c$, is the amplitude at the lowest drive power at which an infinitesimal change in frequency results in an infinite change in amplitude. The analysis of nonlinear response curves and the onset of nonlinearity can be found in \cite{Landau1960}. According to \cite{Yurke1995} the critical amplitude, which is not at the resonance frequency, is given by:
\begin{equation}
	\mathcal{A}_c=\frac{2\cdot 0.83}{3}\omega_0\sqrt{\frac{2\sqrt{3}}{\alpha Q}},
	\label{eq:mathcalAc}
\end{equation}
where $\alpha$ is the nonlinearity parameter which will be discussed in the next section. 

The dynamic range is defined as the ratio of the 1 dB compression point to the noise amplitude at the resonance frequency:
\begin{eqnarray}
	DR(\textrm{dB})&=&20 \,\textrm{log}\!\left(\frac{0.745\mathcal{A}_c}{\sqrt{S_u\Delta\! f}}\right) \nonumber \\
	&=&20 \,\textrm{log}\!\left(\frac{0.745\cdot0.83\,\omega_0^2}{3 Q}\sqrt{\frac{2\sqrt{3}m \omega_0}{k_bT\alpha\Delta\! f}}\right),
	\label{eq:DR}
\end{eqnarray}
where $\Delta \!f$ is the measurement bandwidth. The 1 dB compression point occurs when the peak amplitude is 1 dB lower than what would be expected for a purely linearly response and has a value of $0.745\mathcal{A}_c$. Equation \ref{eq:DR} shows that the low mass and the high nonlinearity parameter for CNT resonators give rise to a small dynamic range compared to top-down fabricated nanomechanical resonators.

\subsection{Geometric nonlinearity}
\label{subs:geom}
In this subsection, it is described how geometric nonlinearities in the motion of CNT resonators arise. When the amplitude of oscillation becomes larger, the CNT experiences significantly more tension at the extremal points of oscillation than at the equilibrium position. This tension thus depends on displacement and is one contribution to the geometric nonlinearity. Furthermore, the force induced by the gate voltage causes tension in the CNT, which again depends on displacement, being another contribution to the geometric nonlinearity.

\begin{figure}[h]
\centering
	\includegraphics[scale=0.60]{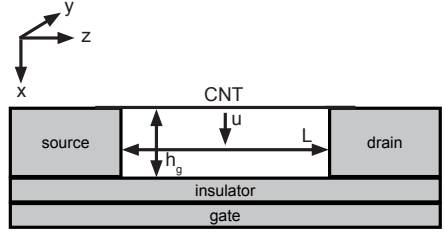}
	\caption{Coordinate axes: the CNT resonator of length $L$ is suspended between the source and drain electrodes at a distance of $h_g$ above the gate. The CNT is aligned in the z-direction. The displacement $u$ takes place in the x-direction.}
	\label{fig:Coordinateaxes}
\end{figure}

To derive the expressions describing geometric nonlinearity we start with the theoretical framework for a driven damped beam \cite{Poot2011}. The Euler-Bernouilli beam theory describes the motion of continuous bodies subjected to external forces. Even though the diameter of a CNT consists of only around 10 atoms, continuum mechanics is nonetheless a good description of the mechanical properties of a CNT resonator \cite{Desquesnes2002}. We take the coordinate system as shown in figure \ref{fig:Coordinateaxes}: The CNT axis is positioned in the z-direction and the oscillation takes (at least for now) place in the x-direction. 
The displacement $u$ of the CNT resonator as a function of time is governed by the following equation of motion:
\begin{equation}
\rho A \frac{\partial^2 u}{\partial t^2}+ \eta \frac{\partial u}{\partial t}+D \frac{\partial^4 u}{\partial z^4}-T \frac{\partial^2 u}{\partial z^2}=f.	
\label{eq:eqmotunscaled}
\end{equation}
The first term describes inertia, with $\rho$ being the density and $A$ the area of the cross section. The second term describes damping with $\eta=\rho A \omega_0 / Q$ the damping rate per unit of length. The third term is the restoring force due to the bending rigidity $D=EI/(1-\nu^2)$, with $E=1.3$ TPa \cite{Krishnan1998} and $\nu =0.2$ \cite{Popov2000} the Young's modulus and Poisson's ratio respectively, and the second (bending) moment of inertia $I$. The CNT's Poisson's ratio is neglected and the denominator is set to unity. We approximate the CNT resonator as a solid cylinder with a second moment of inertia $I = \pi r^4/4$. The fourth term describes the restoring force due to the tension $T$ and $f$ describes an external force per unit length.

There are four boundary conditions determining the modeshapes $u(z)$. The CNT is doubly clamped, leading to $u(0)=0$ and $u(L)=0$. Furthermore, the bending rigidity of the CNT prescribes the slope of the CNT to be continuous at the clamping points, as the CNT goes from lying on the contact to being suspended, resulting in $\partial u/\partial z(0)=0$ and $\partial u/\partial z(L)=0$.

To solve equation \ref{eq:eqmotunscaled} we first separate all contributions into a static/dc and a dynamic/ac component. For the position dependent displacement of the CNT, $u(z)$, and for the gate voltage, $V_g(z)$, this gives:
\begin{eqnarray}
u(z)& = &u_{dc}(z)+u_{ac}(z)\cos(\omega t), \\
V_g(z) &= &V_g^{dc}(z)+V_g^{ac}(z)\cos(\omega t).
\label{eq:}
\end{eqnarray}
The tension felt by the CNT is then given by:
\begin{eqnarray}
T(z)&=&T_0+ \frac{EA}{2L}\int_0^L{\left(\frac{\partial u}{\partial z}\right)^2}dz\nonumber \\ \nonumber
&= &T_0 + \frac{EA}{2L}\int_0^L{\left(\frac{\partial u_{dc}}{\partial z}\right)^2}dz +\frac{EA}{L}\left(\int_0^L{\frac{\partial u_{dc}}{\partial z}\frac{\partial u_{ac}}{\partial z}}dz\right) \cos(\omega t)\\ 
&=&T_{dc}+T_{ac} \cos(\omega t),
\label{eq:}
\end{eqnarray}
with $T_0$ the residual tension present at zero gate voltage. The uniform force per unit length on the CNT due to the gate voltage is:
\begin{eqnarray}
	f(z)& =& \frac{1}{2}\frac{dc_g}{du}V_g^2 \\ \nonumber
&	=& \frac{1}{2}\frac{dc_g}{du}(V_g^{dc})^2+\frac{dc_g}{du}V_g^{dc}V_g^{ac}\cos(\omega t) \\ \nonumber
&	=& f_{dc}(z) + f_{ac}(z)\cos(\omega t).
\label{eq:f}
\end{eqnarray}
Here $dc_g/du$ is the derivative of the capacitance per unit length with respect to displacement. We arrive at two equations; one for the static and one for the dynamic displacement:
\begin{eqnarray}
D \frac{\partial^4 u_{dc}}{\partial z^4}-T_{dc} \frac{\partial^2 u_{dc}}{\partial z^2}=f_{dc},	\\ 
-\omega^2 \rho A u_{ac}+i\omega \eta u_{ac}+D \frac{\partial^4 u_{ac}}{\partial z^4}-T_{dc} \frac{\partial^2 u_{ac}}{\partial z^2}-T_{ac} \frac{\partial^2 u_{dc}}{\partial z^2}=f_{ac}.	
\label{eq:EB0}
\end{eqnarray}
It is useful to introduce scaled variables denoted with primes:

\begin{equation}
\begin{tabular*}{10cm}{l l l }
	$u'_{dc,ac}=\frac{u_{dc,ac}}{r}$, 
		&$z' =\frac{z}{L}$,
		&$T'_{dc,ac}=\frac{T_{dc,ac}L^2}{D}$, \\
		$L'_{dc,ac}=\frac{L_{dc,ac}}{r}=\frac{L^4f_{dc,ac}}{Dr}$,
		&$t'=t\Omega$,
		&$\omega'=\frac{\omega}{\Omega}$,\\
		$\eta'=\frac{\eta L^4\Omega}{D}$,
		&$\Omega = \frac{1}{L^2}\sqrt{\frac{D}{\rho A}}$.&
\end{tabular*}
\end{equation}

The equation for the dynamic displacement now becomes:
\begin{equation}
	-\omega'^2 u'_{ac}+i\omega'\eta' u'_{ac}+D \frac{\partial^4 u'_{ac}}{\partial z'^4}-T'_{dc} \frac{\partial^2 u'_{ac}}{\partial z'^2}-T'_{ac} \frac{\partial^2 u'_{dc}}{\partial z'^2}=L'_{ac}.
	\label{eq:dynmot}
\end{equation}
The ac displacement is separated into an amplitude $\mathcal{A}(t)$ changing slowly with respect to the oscillation, and a spatial part $\xi(z)$:
\begin{equation}
	u_{ac}(z,t)=\mathcal{A}_n(t)\xi_n(z),
	\label{eq:uac}
\end{equation}
where the subscript $n$ denotes the mode of oscillation. The eigenfunctions of the spatial part of equation \ref{eq:dynmot} give the modeshapes:
\begin{equation}
	\mathcal{L}\xi_n(z')=\left(D \frac{\partial^4 }{\partial z'^4}-T'_{dc} \frac{\partial^2}{\partial z'^2}\right)\xi_n-T'_{ac}[\xi_n] \frac{\partial^2 u'_{dc}}{\partial z'^2}=\omega'^2_n(T'_{dc},u'_{dc})\cdot \xi_n(z'),
\end{equation}
where the eigenfrequencies are given by:
\begin{equation}
	\omega'_n=\beta_n^2\sqrt{1+\frac{T'_{dc}}{\beta_n^2}},
	\label{eq:omegan}
\end{equation}
with $\beta_n$ a numerical factor. The first four $\beta_n$ are found to be 4.73004, 7.8532, 10.9956, and 14.1372. Like a guitar string, the resonance frequency of the CNT increases with increased tension. Also, as tension is increased the resonance frequencies converge towards a harmonic spectrum.

In order to make the amplitude $\mathcal{A}_n(t)$ of the symmetric, odd modes correspond to the root-mean-square displacement over the length of the CNT (not in time) as:
\begin{equation}
	\mathcal{A}_n(t)=\sqrt{\frac{\int_0^L{u_{ac}^2dz}}{L}},
	\label{eq:mathcalA}
\end{equation}
the eigenfunctions $\xi_n(x')$ have the following orthonormalization:
\begin{equation}
	\int_0^1{\xi_m(z')\xi_n(z')dz'}=\delta_{mn}.
\end{equation}
Inserting expression \ref{eq:mathcalA} into equation \ref{eq:dynmot}, multiplying by $\xi_n(z')$, integrating over the length, and unscaling gives:
\begin{equation}
	\left(\omega_n^2-\omega^2+i\frac{\omega\omega_n}{Q}\right)\mathcal{A}_n=\frac{f_{ac}}{\rho A}a_n,
\end{equation}
where $a_n=\int_0^1{\xi_n(z')dz'}$. The first four $a_n$ are: 0.83, 0, 0.36, 0. In the ideal case, even modes have a displacement that on average gives no change in capacitance and are therefore often not visible in experiments. 

Up to this point, a linear system was described. We now introduce nonlinearity, looking at two different contributions to the geometric nonlinearity of a CNT resonator. For the first contribution, the CNT is described as a beam \cite{Cleland2003,Nayfeh2000}, with no dc displacement or tension $u_{dc},T_{dc}=0$. For the second contribution, the two regimes from \cite{Sapmaz2003}, i.e. the weak-bending regime and the strong-bending regime, are used to account for the dc displacement a CNT possesses and the tension it acquires from applying a gate voltage.

Modeling the CNT as a beam, we see that when the amplitude of oscillation becomes larger, it is significantly longer at the extremal points of oscillation than at its equilibrium position. A segment that used to be of length $dL=dz$ now has a length of $dL=\sqrt{dz^2+du^2}\approx dz+\frac{1}{2}(\frac{\partial u}{\partial z})^2 dz$. The CNT becomes longer by an amount $\Delta L$ and this causes a change in tension of $EA\Delta L/L$. The associated added tension $T'^*_{ac}$ is:
\begin{eqnarray}
	T'^*_{ac}&= &2\int_0^1{\left(\frac{\partial u'_{ac}}{\partial z'}\right)^2dz'}\nonumber\\
	&=&2\frac{\mathcal{A}_n^2}{r^2}b_n,
	\label{eq:Taccastac}
\end{eqnarray}
where $b_n=\int_0^1{(\frac{\partial \xi_n(z')}{\partial z'})^2dz'}$, with the first four $b_n$ calculated to be 12.3, 46.1, 98.9, and 171.6.
The scaled restoring force per unit length due to the tension $T'^*_{ac}$ is given by:
\begin{equation}
	-f_r'^*=-T'^*_{ac}\frac{\partial^2 u'_{ac}}{\partial z'^2},
	\label{eq:frdyn}
\end{equation}
which we add to equation \ref{eq:dynmot}. Multiplying the augmented equation by $\xi_n(z')$, integrating over the length and unscaling, yields the well-known Duffing equation:
\begin{equation}
	\frac{\partial^2\mathcal{A}_n}{\partial t^2}+\frac{\omega_n}{Q}\frac{\partial\mathcal{A}_n}{\partial t}+\omega_n^2\mathcal{A}_n+\alpha_{beam}c_n\mathcal{A}_n^3=\frac{f_{ac}}{\rho A}a_n,
\end{equation}
with
\begin{equation}
	\alpha_{beam}=\frac{b_n^2E}{2L^4\rho c_n}.
	\label{eq:alphageom}
	\end{equation}
Here the identity $\int_0^1{\frac{\partial^2 \xi_n}{\partial z^2}\xi_n dz}=-\int_0^1{(\frac{\partial \xi_n}{\partial z})^2dz=-b_n}$ is used. The factor \linebreak $c_n=\int_0^1\xi_n(z')^4 dz'$ arises from our definition of $\alpha$: if the analysis had been started not with the unscaled version of equation \ref{eq:frdyn} but with $f_rL=-m\alpha_{beam} u^3$ the Duffing term would have become $\alpha_{beam}\left(\int_0^1\xi_n(z')^4 dz'\right) u^3$. The first four $c_n$ values are 1.85, 1.69, 1.64, and 1.61.From equation \ref{eq:alphageom} it is clear that if the CNT resonator is modeled as a beam the nonlinearity is positive and thus always gives a stiffening spring behaviour. For our example CNT in table \ref{tab:prop}, it results in a nonlinearity for the fundamental mode of $9.6\cdot 10^{34}$ N/(kg m$^3$). 

Until now, we have assumed that the CNT acts as a beam with no static displacement and no static tension. In CNT devices, these assumptions are often not valid: by applying a dc voltage to a nearby gate, the CNT can acquire significant static displacements that can be larger than the CNT diameter. As described in \cite{Sapmaz2003}, with no initial tension, there are two regimes describing the bending of the CNT: the weak-bending and the strong-bending regime. In the weak-bending regime, the CNT acts as a beam at low gate voltages, and both $\alpha_{beam}$ and $\alpha_{weak}$ contribute to the geometric nonlinearity. It undergoes a transition to string-like behaviour at larger gate voltages as the CNT becomes strongly bent by the force from the gate. The crossover from the weak-bending to the strong-bending regime occurs when the restoring force due to tension is equal to the restoring force due to bending rigidity, $T_{dc}=EI/L^2$, or equivalently, when the CNT displaces by a distance equal to its radius. At this point the contribution $\alpha_{beam}$ to the geometric nonlinearity, arising from the tension induced by the oscillation of the CNT, can no longer be described by equation \ref{eq:alphageom}. This is because, at non-zero dc deflection in the strong-bending regime, the oscillation of the CNT does not lead to increased tension when moving away from the gate. The gate voltage $V_g^{*}$ at which the crossover from the weak-bending regime to the strong-bending regime occurs is found from $l'_{dc}\approx 871$ and for the example CNT in table 1 has a value of $V_g^{*}=2.1$V.

At low gate voltages, the CNT is in the weak-bending regime with $T_{dc}\ll EI/L^2$. In this range of gate voltages, the restoring force is dominated by the bending rigidity, leading to mechanical modes with resonance frequencies at ratios close to that of an ideal beam. Starting from a straight CNT (i.e. no dc deflection at zero gate voltage), the tension induced by the gate is \cite{Sapmaz2003}:
\begin{equation}
T_{dc}^{weak}=\frac{L^4A}{60480EI^2}F_{dc}^2,
\label{eq:Tweakbending}
\end{equation}
and the mechanical resonance frequency of the lowest mode is given by:
\begin{eqnarray}
\omega_0^{weak}&=&\sqrt{\frac{EI}{\rho A}}\frac{\beta_0^2}{L^2}+0.28\frac{T}{\sqrt{\rho AEI}}\nonumber\\
&=&\sqrt{\frac{EI}{\rho A}}\frac{\beta_0^2}{L^2}+0.07\frac{L^4 A}{60480 EI^2\sqrt{\rho AEI}}\left(\frac{dC_g}{du}\right)^2 V_g^4.
\label{eq:omeganweak}
\end{eqnarray}
For a CNT that begins with no deflection, the resonance frequency at low gate voltage in this beam-like, weak-bending regime scales with $\omega \propto V_g^4$. Note that this quartic gate voltage dependence arises due to the fact that the electrostatic force from the gate is perpendicular to the axis of the CNT, and therefore does not begin to add tension until the CNT develops a static displacement. If the CNT already has a static displacement at zero gate voltage, due, for example, to buckling from compressive strain, or from a downwards curvature of the substrate at the edge of the trench, then the resonance frequency in the weak-bending regime shows a $\omega \propto V_g^2$ gate dependence. From expression \ref{eq:omeganweak} for the resonance frequency, we obtain the gate voltage dependent spring constant:
\begin{equation}
k_{weak}=\frac{EI}{\rho A}\frac{\beta_0^4}{L^4}+0.14\frac{L^3 A \beta_0^2}{60480 EI^2}\left(\frac{dC_g}{du}\right)^2 V_g^4.
\label{eq:kweak}
\end{equation}
We obtain the Duffing nonlinearity parameter by taking the second derivative of equation \ref{eq:kweak} with respect to displacement. We do this by using the fact that the displacement of the CNT resonator leads to a change in capacitance to the gate. This in turn is interpreted as an effective change in the gate voltage, since only the gate-induced charge $q_c=C_gV_g$ is relevant.
\begin{eqnarray}
\alpha_{weak}&=&\frac{1}{2m}\frac{d^2 k}{d u^2}=\frac{1}{2m}\left(\frac{V_g}{C_g}\frac{d C_g}{d u}\right)^2\frac{d^2 k}{d V_g^2}\nonumber\\
&=&0.84\frac{L^2\beta_0^2}{60480 E I^2C_g^2 \rho}\left(\frac{d C_g}{d u}\right)^4V_g^4.
\label{eq:}
\end{eqnarray}
For the example CNT in table \ref{tab:prop} this results in a nonlinearity of $8.7\cdot 10^{28}\cdot (V_g \textrm{ in Volt})^4$ N/(kg m$^3$). Note that this gate-induced contribution to the nonlinearity in the weak- bending regime is much smaller than the contribution from the intrinsic gate-voltage independent beam nonlinearity (equation \ref{eq:alphageom}), and thus the nonlinearity in the weak-bending regime is dominated by that described in equation \ref{eq:alphageom}.

As the gate voltage is increased further, the CNT enters the strong-bending regime, in which its static displacement exceeds its diameter, and the restoring force from the induced tension exceeds the bending rigidity. In this regime, the mechanical modes begin to behave as those of a string under tension, approaching a harmonic spectrum at large gate voltages. In this regime, the gate induced tension is given by \cite{Sapmaz2003}:
\begin{equation}
T_{dc}^{strong}=\left(\frac{EA}{24}\right)^\frac{1}{3}F_{dc}^\frac{2}{3},
\label{eq:}
\end{equation}
with a resonance frequency for the lowest mode given by:
\begin{equation}
\omega_0^{strong}=\frac{\pi}{L}\sqrt{\frac{T_{dc}^{strong}}{\rho A}}=\frac{\pi}{L\sqrt{\rho A}}\left(\sqrt{\frac{EA}{96}}\frac{dC_g}{du}\right)^\frac{1}{3}V_g^\frac{2}{3}.
\label{eq:omeganstrong}
\end{equation}
In this strong-bending regime, the resonance frequency scales as $\omega \propto V_g^{\frac{2}{3}}$. The gate voltage dependent spring constant is given by:
\begin{equation}
k_{strong}=\frac{\pi^2}{L}\left(\frac{EA}{96}\right)^\frac{1}{3}\left(\frac{dC_g}{du}\right)^\frac{2}{3}V_g^\frac{4}{3},
\label{eq:kstrong}
\end{equation}
and we obtain the following for the Duffing nonlinearity parameter:
\begin{equation}
\alpha_{strong}=\frac{2\pi^2}{9mLC_g^2}\left(\frac{d C_g}{d u}\right)^\frac{8}{3} \left(\frac{EA}{96}\right)^\frac{1}{3} V_g^\frac{4}{3}.
\end{equation}
For the example CNT in table \ref{tab:prop} this results in a nonlinearity of $1.9\cdot 10^{28}\cdot (V_g \textrm{ in Volt})^\frac{4}{3}$ N/(kg m$^3$). 

\begin{figure}[h]
\centering
	\includegraphics[scale=0.70]{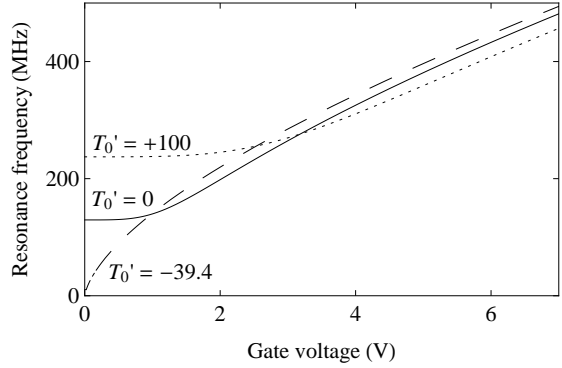}
	\caption{Simulation showing the resonance frequency as a function of gate voltage for three values of residual tension $T_0'$. At a high positive residual tension of $T_0'=100$ (dotted line), the resonance frequency only starts to increase when the tension due to the gate voltage is similar to the residual tension. At zero residual tension $T_0'=0$ (solid line), the tension due to the gate voltage first has to overcome the bending rigidity in the weak-bending regime with $\omega_0 \propto V_g^2$, and then the strong-bending regime with $\omega_0 \propto V_g^{\frac{2}{3}}$ is entered. At a negative residual tension $T_0'=-39.4$ signifying residual compression (dashed line), the resonance frequency increases significantly with gate voltage even at low gate voltages.}
	\label{fig:resfreqs}
\end{figure}

Until now, we have assumed that the CNT includes no initial tension, i.e. that its relaxed length is the same as the length of the trench. For positive initial tension, corresponding to a tensile stress, the CNT behaves already as a string at zero gate voltage, and shows a flat resonance frequency vs. gate response. This flat gate response of the resonance frequency continues until the gate induced tension exceeds this initial tension and begins to tune the resonance frequency, as shown in figure \ref{fig:resfreqs}. 

If the CNT relaxed length is longer than the distance across the trench, the CNT experiences a compressive force. For compressive forces below the Euler instability, this results in a reduction of the resonance frequency at low gate voltages. As the compressive force approaches the Euler instability, the resonance frequency at zero gate voltage vanishes, as shown in figure \ref{fig:resfreqs}. For compressive forces beyond the Euler instability, the CNT buckles and acquires a non-zero static displacement even at zero gate voltage. Beyond the buckling point, the resonance frequency becomes finite again, and the CNT exhibits bending-like modes around the buckled beam shape. In this case, the bending modes show an $\omega \propto V_g^2$ gate dependence due to the non-zero initial displacement, as discussed above. The modes of such buckled beams can display peculiar mechanical modes similar to those of a hanging swinging chain. A complete analysis of the mechanical modes of strongly buckled CNTs can be found in \cite{Ustunel2005}.

In the strong bending regime, the gate voltage dependent spring constant given by equation \ref{eq:kstrong} results in a significant quadratic contribution to the restoring force. This suggests that the quadratic renormalization of the cubic force coefficient may play an important role in this case, something that should be the subject of further investigation.

%

\subsection{Electrostatic nonlinearity}

A second origine of nonlinearity comes from the electrostatic interaction between the CNT and the gate electrode. If for a moment we consider the CNT not to experience Coulomb blockade, but to be a perfect conductor, then it screens the voltage applied to the gate. The electrostatic force between the CNT and the gate was already given in equation \ref{eq:f} as:
\begin{equation}
F_{ES}=- \frac{dU}{du} = \frac{1}{2} \frac{dC_g}{du} V_g^2.
\label{eq:}
\end{equation}
In the case of a cylinder above a plate, the capacitance per unit length $c_g=C_g/L$ is given by \cite{Sapmaz2003}:
\begin{equation}
c_g(z)=\frac{2 \pi \epsilon_0}{\textrm{arccosh}\!\left(\frac{h_g-u(z)}{r}\right)},
\label{eq:}
\end{equation}
where $h_g$ is the distance between the CNT and the gate. For the example CNT in table \ref{tab:prop} this gives values of $5.9\cdot 10^{-21} \textrm{ F/nm}$, $3.5\cdot 10^{-23} \textrm{ F/nm}^2$, and $2.9\cdot 10^{-27} \textrm{ F/nm}^4$ for the first, second, and fourth derivative of the capacitance to gate with respect to displacement respectively.
The force acting on the CNT due to electrostatics can be expanded as:
\begin{equation}
	F_{ES}=-\Delta k_{ES} u=-\left(\Delta k_{ES,0}+m\alpha_{ES} u^2+\mathcal{O}(u^4)\right)u,
\end{equation}
and adds to the spring constant as:
 \begin{eqnarray}
	\Delta k_{ES} &=&- \frac{dF_{ES}}{du} \nonumber \\
	&=&-\frac{1}{2}\frac{d^2C_g}{du^2}V_g^2.
\end{eqnarray}
The electrostatic force thus gives a contribution to the spring constant that is opposite in sign to the contribution from the restoring force due to bending rigidity and tension. Note, in particular, that the electrostatic forces do {\em not} act like a restoring force: for example, if the CNT is displaced away from its equilibrium position towards the gate, the electrostatic force pulling it towards the gate {\em increases}. This has the effect of an ``anti''-restoring force. In the case that this electrostatic ``anti''-restoring force is larger than the restoring force from the mechanical deformation, the system can even become unstable, leading to a phenomenon of electrostatic ``snap in'', in which the device is pulled uncontrollably towards the gate until it makes physical contact. 

In the case that the restoring force from mechanical deformation dominates, the system is stable, but the effective spring constant is reduced by the electrostatic interaction. The reduction in the spring constant leads to a decrease in resonance frequency with gate voltage. When the CNT is close enough to the gate so that $d^2C_g/du^2$ is large, this decrease in resonance frequency is larger than the increase in resonance frequency with gate voltage due to increased tension.

The nonlinearity parameter due to the electrostatic force is:
\begin{eqnarray}
	\alpha_{ES}&=&-\frac{1}{6m}\frac{d^3F_{ES}}{du^3} \nonumber \\
	&=&-\frac{1}{12m}\frac{d^4C_g}{du^4}V_g^2.
\end{eqnarray}

The electrostatic nonlinearity parameter is negative and always gives rise to a softening spring nonlinearity.
For the example CNT in table \ref{tab:prop}, the nonlinearity due to electrostatics is $-3.2 \cdot 10^{28}\cdot (V_g \textrm{ in Volt})^2$ N/(kg m$^3$). This is of the same order of magnitude as the geometric nonlinearities $\alpha_{weak}$ and $\alpha_{strong}$, but has an opposite sign.

\subsection{Nonlinearity due to single-electron tunneling}
\begin{figure}[h]
\centering
	\includegraphics[scale=0.70]{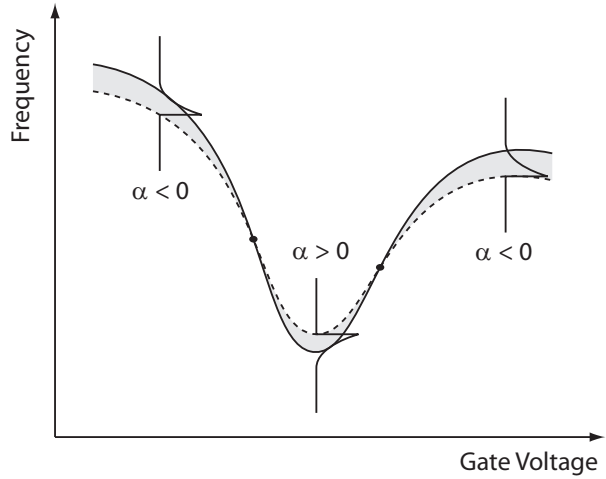}
	\caption{Comparison of two effects caused by single-electron tunneling: dips in the mechanical frequency, and mechanical nonlinearity. When the curvature of the change in resonance frequency is negative, so is the nonlinearity parameter. A positive curvature of the change in resonance frequency leads to a positive nonlinearity parameter.}
	\label{fig:Alphavscurvature}
\end{figure}

In the previous section we described how single-electron tunneling causes a softening spring nonlinearity. Here, we describe how it is also an additional source of nonlinearity. Writing the force on the CNT due to single electron tunneling as:
\begin{equation}
	F_{SET}=-\Delta k_{SET} u=-\left(\Delta k_{SET,0}+m\alpha_{SET} u^2+\mathcal{O}(u^4)\right)u,
\end{equation}
where $\Delta k_{SET}$ is given by equation \ref{eq:deltak}, gives for the nonlinearity due to single-electron tunneling:
\begin{equation}
\alpha_{SET}=\frac{1}{6m}\frac{d^3F_{SET}}{du^3}=\frac{1}{2m}\frac{d^2\Delta k_{SET}}{du^2}=\frac{1}{2m}\left(\frac{V_g}{C_g}\frac{dC_g}{du}\right)^2\frac{d^2\Delta k_{SET}}{dV_g^2}.
	\label{eq:alphaSET1}
\end{equation}
We now use the expression from equation \ref{eq:deltak} and insert it into equation \ref{eq:alphaSET1}. Assuming that the gate voltage difference in which the charge transition occurs is much smaller than the gate voltage, and neglecting terms of $d^2C_g/du^2$, we arrive at:
\begin{equation}
	\alpha_{SET}=-\frac{e}{2m}\frac{V_g^3(V_g-V_{CNT})}{C_g^3C_{tot}}\left(\frac{dC_g}{du}\right)^4\frac{d^3\!\left\langle N \right\rangle}{dV_g^3}.
	\label{eq:alphaSET2}
\end{equation}
Equation \ref{eq:alphaSET1} shows that the nonlinearity parameter is proportional to the curvature of the change in spring constant and consequently to the curvature of the change in resonance frequency. Figure \ref{fig:Alphavscurvature} illustrates how the sign of $\alpha_{SET}$ changes with gate voltage: in the middle of the frequency dip caused by single-electron tunneling the curvature of the change in resonance frequency is positive, leading to a positive nonlinearity, whereas on the left and right side of the frequency dip, both the curvature and the nonlinearity are negative.

\begin{figure}[h]
\centering
	\includegraphics[scale=0.70]{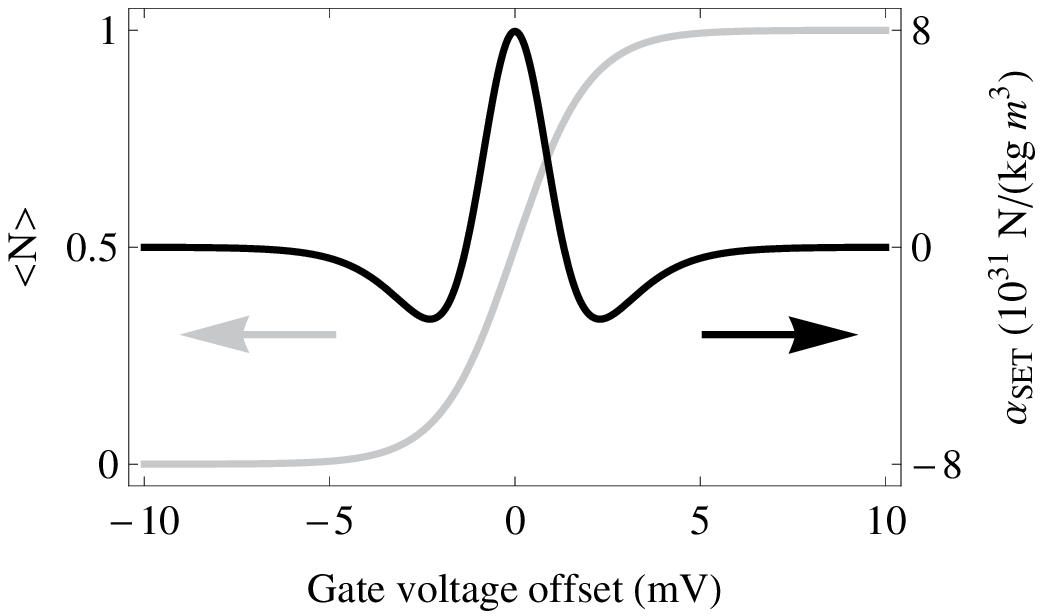}
	\caption{Calculated average occupancy (left axis) and the nonlinearity parameter $\alpha_{SET}$ due to single-electron tunneling (right axis): $\alpha_{SET}$ depends on the third derivative of the average occupancy with respect to gate voltage, thus going from negative to positive to negative again in only a few mV.}
	\label{fig:AlphaSET}
\end{figure}

According to equation \ref{eq:alphaSET2} the nonlinearity parameter depends on the third derivative of the average occupancy with respect to gate voltage. Figure \ref{fig:AlphaSET} shows, using the values from the example CNT in table \ref{tab:prop}, how by changing the gate voltage a few mV, the nonlinearity parameter changes dramatically and even changes sign. For $V_{CNT}\ll V_g$ and a charge transition that occurs in $\Delta V_g=10$ mV the example CNT has a minimum (i.e. most negative) nonlinearity due to single-electron tunneling of $\alpha_{SET}=-2.7 \cdot 10^{31} \cdot (V_g \textrm{ in Volt})^4 $N/(kg m$^3)$. The maximum (i.e. most positive) nonlinearity is $\alpha_{SET}=+8.0 \cdot 10^{31} \cdot (V_g \textrm{ in Volt})^4 $ N/(kg m$^3)$. At the inflection point of the frequency dip, the cubic, Duffing nonlinearity due to single-electron tunneling vanishes. At this point other nonlinearities, such as the quadratic nonlinearity, will dominate the nonlinear dynamics, which suggest further theoretical and experimental investigation of this regime.

In table \ref{tab:Comparealphas} the magnitude of the four different contributions to the nonlinearity parameter for the example CNT in table \ref{tab:prop} is depicted. The table demonstrates that the nonlinearity due to single-electron tunneling, $\alpha_{SET}$, is similar in size to the beam-like geometric nonlinearity, $\alpha_{beam}$. Which nonlinearity dominates depends on the gate voltage and the gate voltage range over which the charge transition occurs. The nonlinearities due to weak-bending and strong-bending and the electrostatic nonlinearity have almost the same value, but opposite sign, and can be neglected for a quantum dot CNT resonator.

\begin{table*}[h]
	\begin{tabular}{|l|c|c|c|}
		\hline
				Nonlinearity (N/(kg m$^3$))&$V_g= 1$ V&$V_g= V_g^*=2.1$ V&$V_g= 5$ V\\
		\hline
			$\alpha_{SET,max}$ ($\Delta V_g=10$ mV)&$8.0\cdot 10^{31}$ &$1.5\cdot 10^{33}$ &$5.0\cdot 10^{34}$ \\
			$\alpha_{SET,min}$ ($\Delta V_g=10$ mV)&$-2.7\cdot 10^{31}$ &$-5.2\cdot 10^{32}$&$-1.7\cdot 10^{34}$ \\
			$\alpha_{SET,max}$ ($\Delta V_g=1$ mV)&$8.0\cdot 10^{34}$ &$1.5\cdot 10^{36}$ &$5.0\cdot 10^{37}$ \\
			$\alpha_{SET,min}$ ($\Delta V_g=1$ mV)&$-2.7\cdot 10^{34}$ &$-5.2\cdot 10^{35}$ &$-1.7\cdot 10^{37}$ \\
			$\alpha_{ES}$&$-3.2\cdot 10^{28}$ & $-1.4\cdot 10^{29}$&$-7.9\cdot 10^{29}$ \\
			$\alpha_{weak}$&$8.7\cdot 10^{28}$ & $1.7\cdot 10^{30}$&- \\
			$\alpha_{strong}$&- &$5.0\cdot 10^{28}$ &$1.6\cdot 10^{29}$ \\
			$\alpha_{beam}$&$9.6\cdot 10^{34}$ &$9.6\cdot 10^{34}$ &- \\
			\hline
		 
		 \end{tabular}
		\caption{Comparison of the magnitude of the parameters for the four contributions to the nonlinearity parameter for the example CNT in table \ref{tab:prop}. For the nonlinearities due to single-electron tunneling the gate voltage range $\Delta V_g$ over which the charge transition occurs is indicated. The nonlinearities in the weak-bending, beam-like (strong-bending) regime are not calculated above (below) the crossover voltage $V_g^*$.}
		\label{tab:Comparealphas}
\end{table*}


\section{Parametric excitation and mode coupling in carbon nanotube resonators}

The previous section showed how geometry, electrostatics, and single-electron tunneling give rise to a Duffing-type nonlinearity. In addition to a nonlinear response to large powers at the resonance frequency, nonlinear effects can also be observed when driving the resonator at frequencies different from its resonance frequency. In the next subsection, we look at how nonlinearities in CNTs can result in parametric driving and amplification. In subsection \ref{subs:MC}, we consider other modes of oscillation, either in another dimension of oscillation, giving rise to nonplanar motion, or as higher order modes in the same direction. Through the nonlinear behaviour of the device, these other modes become coupled to the fundamental mode. As we will see, driving these other modes can generate, for example, extra tension in the device, which can then shift the frequency of the fundamental mode.

\subsection{Parametric excitation}

A resonator is parametrically excited when besides a driving force around $\omega_0$ there is also a change in one of its parameters around $2\omega_0$. A well-known example entails a child on a swing which alters its center of mass during swinging. In this analysis we only consider a modulation of the spring constant, so that $k=k_0+k_p\cos(\omega_pt)$. 

The pumping of energy into the motion through parametric excitation is maximum when done at twice the resonance frequency, but can be done at any frequency for which $\omega_p = \frac{2}{n}\omega_0$, becoming less effective for larger $n$. The equation of motion describing parametric excitation is:
\begin{equation}
\frac{\partial^2u}{\partial t^2}+\frac{\omega_0}{Q}\frac{\partial u}{\partial t}+\frac{1}{m}\left(k_0+k_p\cos(\omega_p t)\right)u=\frac{F_{ac}}{m}\cos(\omega t).
\label{eq:}
\end{equation}
In the case of capacitive driving at $\omega_p$, the change in the spring constant can be described as:
\begin{eqnarray}
\Delta k=-\frac{d F}{d u}&=&\frac{1}{2}\frac{d^2C_g}{d u^2}\left((V_g^{dc})^2+2V_g^{dc}V_g^{ac} cos(\omega_p t)+(V_g^{ac})^2\cos(2\omega_p t)\right)\nonumber\\
&=&\Delta k_0+k_p\cos(\omega_pt)+k_p'\cos(2\omega_pt),
\label{eq:kp}
\end{eqnarray}
yielding an extra modulation of the spring constant $k_p'$ at $2\omega_p$. For low drive powers, $V_g^{ac}\ll V_g^{dc}$, so that $k_p'\ll k_p$, and $k_p'$ can be neglected. The most effective parametric excitation can then be achieved by driving at $\omega_p=2\omega_0$. 

Recently, parametric excitation has been applied to a CNT resonator \cite{Eichler2011}. A CNT resonator is especially suited for parametric excitation because the change in spring constant with drive voltage, $dk/dV_g$, is large (see equations \ref{eq:kweak} and \ref{eq:kstrong}).

\subsection{Mode coupling}
\label{subs:MC}
Until now we have only considered oscillations of the fundamental mode taking place in the x-direction. In this section we discuss coupling between the fundamental mode in the x-direction to the fundamental mode in the y-direction \cite{Ho1975,Conley2008,Perisanu2010} and between the fundamental mode in the x-direction and higher modes in the x-direction \cite{Westra2010}. Both types of mode coupling can generically be described by the following two coupled equations:
\begin{eqnarray}
	\rho A \frac{\partial^2 u}{\partial t^2}+ \eta \frac{\partial u}{\partial t}+D \frac{\partial^4 u}{\partial z^4}-
	 (T_{dc}+T_{u}+T_{v}) \frac{\partial^2 u}{\partial z^2}=f_u, \\
	 	\rho A \frac{\partial^2 v}{\partial t^2}+ \eta \frac{\partial v}{\partial t}+D \frac{\partial^4 v}{\partial z^4}-
	 (T_{dc}+T_{u}+T_{v})\frac{\partial^2 v}{\partial z^2}=f_v.
\end{eqnarray}
Analogously to equation \ref{eq:uac}, the displacement of the other mode is denoted by:
\begin{equation}
	v_{ac} = \mathcal{B}(t)\psi(z).
\end{equation}
Here, $\mathcal{B}(t)$ is the slowly varying amplitude of the other mode, root-mean-squared along the CNT, and $\psi(z)$ depicts the modeshape.

The different parts of the tension on the CNT by the oscillations $u$ and $v$ are given by: 
\begin{eqnarray}
	T_{u}&=&\frac{EA}{2L}\int_0^L{\left(\frac{\partial u_{ac}}{\partial z}\cos(\omega_u t)\right)^2}dz ,\\
	T_{v}&=&\frac{EA}{2L}\int_0^L{\left(\frac{\partial v_{ac}}{\partial z}\cos(\omega_v t)\right)^2}dz.
\end{eqnarray}
Here $T_{u}$ causes the nonlinearity for a beam as described in section \ref{sec4}.
 
The tension arising from the oscillation of $v$ is:
\begin{eqnarray}
	T_{v} &=& \frac{EA}{2L}\mathcal{B}^2\left(\frac{1}{2}+\frac{1}{2}\cos(2\omega_v t)\right)\int_0^L{\left(\frac{\partial \psi}{\partial z}\right)^2dz}\nonumber\\
	&=&\frac{EAb_n^v}{4L^2}\mathcal{B}^2+\frac{EAb_n^v}{4L^2}\mathcal{B}^2\cos(2\omega_v t),
	\label{eq:Ty}
\end{eqnarray}
remembering $\int_0^L{\left(\frac{\partial \psi}{\partial z}\right)^2dz}=b_n^v/L$.
This means there is both a contribution to the dc tension and there is change in the ac tension at $2\omega_v$.

Let us first look at the contribution to the dc tension in the weak-bending limit. Using equation \ref{eq:omeganweak} and \ref{eq:Ty} in the limit of $T_v\ll EI/L^2$ the change in resonance frequency 
 to mode coupling is:
\begin{equation}
	\Delta \omega_0^{weak} =\sqrt{\frac{EA}{\rho I}}\frac{b_n^v}{8 L^2}\mathcal{B}^2.
\end{equation}

For CNTs, the high Young's modulus, low mass density, high ratio of $A/I=4/r^2$, and small length cause a large change in resonance frequency due to nonplanar motion. For the example CNT from table \ref{tab:prop} this gives a tuning of $16$  MHz/nm$^2$ when the other mode is the fundamental mode in the y-direction. For the third mode the tuning of the fundamental mode for the example CNT is $127$ MHz/nm$^2$, arising from a higher $b_n^v$.

In the strong-bending limit with $T_v\ll T_{dc}^{strong}$, using equation \ref{eq:omeganstrong} leads to the change in resonance frequency due to mode coupling:
\begin{eqnarray}
	\Delta \omega_0^{strong} &=&\frac{\pi}{2L\sqrt{\rho A}}\frac{T_v}{\sqrt{T_{dc}^{strong}}}\nonumber\\
	&=&\frac{\pi (E A)^\frac{5}{6}b_n^v}{4\cdot 2^\frac{1}{6}L^3}\left(\frac{d C_g}{d u}\right)^{-\frac{1}{3}}V_g^{-\frac{2}{3}}\mathcal{B}^2.
\end{eqnarray}
As the tension due to the gate electrode increases the added tension from mode coupling is less important. At $V_g=5$V the tuning of the fundamental mode of the example CNT due to another fundamental mode and due to a third mode is reduced to $3.3$ MHz/nm$^2$ and $26$ MHz/nm$^2$ respectively.

\section{Conclusions}
Carbon nanotubes display a rich variety in nonlinear effects. Because of the importance of the measurement apparatus on nonlinear effects, we started this chapter by displaying the multitude of methods to detect the flexural motion of a CNT resonator. In the next section, we illustrated how single-electron tunneling in quantum dot CNT resonators, combined with their low mass, leads to pronounced softening-spring behaviour and significant damping. In section four, we covered the four different contributions to the nonlinear oscillation of a CNT resonator: the beam-like nonlinearity, the nonlinearity due to gate-induced tension, the electrostatic nonlinearity, and the nonlinearity due to single-electron tunneling, the combination of which is unique to CNT resonators. In the final section we show how the large response of the resonance frequency of a CNT resonator to a change in gate voltage or tension makes CNT resonators ideally suited for parametric excitation and mode coupling.

Acknowledgments: We thank Hidde Westra, Giorgi Labadze, Yaroslav Blanter, and Menno Poot for fruitful discussions and we acknowledge the financial support of the Future and Emerging Technologies programme of the European Commission, under the FET-Open project QNEMS (233992).


\end{document}